\input harvmac.tex
\input psfig.sty
\input tables

\def\tilde{\widetilde}
\def\CM{{\cal M}}
\def\lfm#1{\medskip\noindent\item{#1}}
\lref\sen{A. Sen, ``$F$-Theory And Orientifolds,'' hep-th/9605150.}
\lref\bds{T. Banks, M. Douglas, and N. Seiberg, 
``Probing $F$-Theory With Branes,'' hep-th/9605199.}
\lref\ns{N. Seiberg, ``IR Dynamics on Branes and
Space-Time Geometry,'' hep-th/9606017.}
\lref\sw{N. Seiberg and E. Witten, ``Gauge Dynamics and
Compactification to Three Dimensions,'' hep-th/9607163.}
\lref\swoi{N. Seiberg and E. Witten, ``Electric-Magnetic Duality,
Monopole Condensation, And Confinement In $N=2$ Supersymmetric
Yang-Mills Theory,'' Nucl. Phys. {\bf B426} (1994) 19, hep-th/9407087.}
\lref\swoii{N. Seiberg and E. Witten, ``Monopoles, Duality,
And Chiral Symmetry Breaking In $N=2$ Supersymmetric QCD,''
Nucl. Phys. {\bf B431} (1994)  484,  hep-th/9408099.}
\lref\nonren{N. Seiberg, ``Naturalness Versus Supersymmetric
Non-renormalization Theorems,'' Phys.Lett. {\bf B318} (1993) 469,
hep-ph/9309335.}
\lref\aps{P.C. Argyres, M.R. Plesser and N. Seiberg,
``The Moduli Space of Vacua of $N=2$ SUSY QCD and Duality in $N=1$ SUSY
QCD,'' hep-th/9603042.}
\lref\nso{N. Seiberg, ``Electric-Magnetic Duality in Supersymmetric
Non-Abelian Gauge Theories,'' Nucl.Phys. {\bf B435} (1995) 129,
hep-th/9411149.} 
\lref\mp{D.R. Morrison and M.R. Plesser, ``Towards Mirror Symmetry as
Duality for Two Dimensional Abelian Gauge Theories,'' hep-th/9508107.}
\lref\kron{P.B. Kronheimer, Jour. Differential Geometry, 
{\bf 29} (1989) 665.}
\lref\egh{T.  Eguchi, P. Gilkey, and A. Hanson, Phys. Rep. {\bf 66}
(1980) 213.}
\lref\mrd{M.R. Douglas, ``Branes Within Branes,'' hep-th/9512077.} 
\lref\polya{A.M. Polyakov, ``Quark Confinement And The Topology of Gauge
Groups,'' Nucl. Phys. {\bf B234} (1983) {269}.} 

\Title{hep-th/9607207, RU-96-63, IASSNS-HEP-96/80}
{\vbox{\centerline{Mirror Symmetry in}
\centerline{Three Dimensional Gauge Theories}}}
\bigskip
\centerline{K. Intriligator$^1$ and N. Seiberg$^{2}$}
\vglue .5cm
\centerline{$^1$ School of Natural Sciences}
\centerline{Institute for Advanced Study}
\centerline{Princeton, NJ 08540, USA}
\vglue .3cm
\centerline{$^2$Department of Physics and Astronomy}
\centerline{Rutgers University}
\centerline{Piscataway, NJ 08855-0849, USA}

\medskip
\noindent
We discuss non-trivial fixed points of the renormalization group with
dual descriptions in $N=4$ gauge theories in three dimensions.  This new
duality acts as mirror symmetry, exchanging the Higgs and Coulomb
branches of the theories.  Quantum effects on the Coulomb branch arise
classically on the Higgs branch of the dual theory.  We present examples
of dual theories whose Higgs/Coulomb branch are the ALE spaces and whose
Coulomb/Higgs branches are the moduli space of instantons of the
corresponding $ADE$ gauge group.  In particular, we show that in three
dimensions small $E_8$ instantons in string theory are described by a
local quantum field theory.

\Date{7/96}

\newsec{Introduction}

Motivated by \refs{\sen, \bds}
three dimensional gauge theories with $N=4$ supersymmetry were recently
studied {}from the viewpoint of string theory \ns.  One of the main new
results is a new connection between $ADE$ groups and $ADE$
singularities.  A $U(1)$ gauge theory with $n+1$ electrons has a global
symmetry $SU(n+1) = A_n$.  Its moduli space has two branches: a Higgs
branch which is the same as the moduli space of $A_n$ instantons and a
Coulomb branch which has an $A_n$ singularity.  Similarly, an $SU(2)$
gauge theory with $n$ quarks has a global symmetry $SO(2n)=D_n$.  Its
moduli space also has two branches: a Higgs branch which is the same as
the moduli space of $D_n$ instantons and a Coulomb branch with a $D_n$
singularity. 

These theories were further studied in \sw\ where many more details of
their Coulomb branches were analyzed {}from a field theoretic viewpoint.
Also, new relations between these theories and their four dimensional
counterparts \refs{\swoi,\swoii}\ were uncovered.  These theories are
thus a natural setup for studying various physical phenomena and
relations between mathematics and physics.  In this note we continue the
investigation of these theories and find a surprising duality, further
strengthening the link between $ADE$ groups and $ADE$ singularities.

$N=4$ theories in three dimensions have a global $SO(4)\cong
SU(2)_L\times SU(2)_R$, with $SU(2)_R$ the $R$ symmetry seen in six
dimensional $N=1$ theories and $SU(2)_L$ (denoted $SU(2)_{R2}$ in \ns\
and $SU(2)_N$ in \sw) associated with rotations in the three directions
reduced in going to three dimensions.  The Higgs branch is a
hyper-Kahler manifold with Kahler forms transforming in the adjoint of
$SU(2)_R$ and invariant under $SU(2)_L$.  The Coulomb branch is also a
hyper-Kahler manifold with Kahler forms transforming in the adjoint of
$SU(2)_L$ and invariant under $SU(2)_R$.

As in \refs{\nonren,\aps}, we can easily prove a few non-renormalization
theorems by promoting some of the coupling constants to background
superfields: 

\lfm{1.}  When the gauge coupling constant is promoted to a superfield,
the scalars in that superfield transform as $({\bf 1+3,1})$ of
$SU(2)_L\times SU(2)_R$.  Therefore, they can only appear in the Coulomb
branch.  Hence, the Higgs branch is not renormalized by quantum effects
\aps.

\lfm{2.}  Mass terms are background vector fields \refs{\aps, \sw} which
transform as $({\bf 3,1})$ of $SU(2)_L\times SU(2)_R$.  As with the
gauge couplings, they affect the metric on the Coulomb branch but not
the Higgs branch.

\lfm{3.}  Fayet-Iliopoulos $D$ terms for $U(1)$ factors of the gauge
group transform as $({\bf 1, 3})$ of $SU(2)_L\times SU(2)_R$.
Therefore, they can affect only the metric on the Higgs branch but not
the Coulomb branch.

Many of these $N=4$ theories flow in the IR to new non-trivial fixed
points of the renormalization group.  In particular, the theories at the
singularities mentioned above are at such interacting fixed points \ns.
As in four dimensional gauge theories, there could be dual descriptions
of the fixed points \nso.  The duality which we will present exchanges:

\lfm{1.} $SU(2)_L $ and $ SU(2)_R$,

\lfm{2.} The Coulomb branch and the Higgs branch,

\lfm{3.} Mass terms and Fayet-Iliopoulos $D$ terms.

In many respects this duality is reminiscent of the duality of \nso\ in
four dimensional gauge theories and to the presentation of mirror
symmetry in two dimensions given in \mp.  In all these cases the duality
applies only to the long distance theory (the theories at short distance
are free) and the matter fields and gauge fields mix in a non-trivial
way.  Furthermore, as in \mp, terms in the superpotential like mass
terms are exchanged with Fayet-Iliopoulos $D$ terms.

Because the Coulomb branch gets quantum corrections while the Higgs
branch does not, the duality exchange (2.) means that quantum effects in
one theory arise classically in the dual and visa-versa.

Another interesting aspect of the duality is that the fixed point can
have global symmetries which are manifest in one description but arise
quantum-mechanically in the dual.  To see that, note that the mass terms
can be regarded as expectation values of background gauge fields
transforming in the adjoint of any global flavor symmetry; $N=4$
supersymmetry requires them to be in the Cartan subalgebra.
Fayet-Iliopoulos (FI) terms, on the other hand, are not associated with
any visible global symmetry.  Therefore, the exchange (3.) means that
visible global symmetries are exchanged with hidden ones.

Indeed, we can exhibit part of the hidden global symmetries associated
with FI terms.  For every U(1) factor in the gauge group\foot{As found
in \polya, the corresponding conservation law for the non-Abelian part
of the gauge group is violated by instantons.}, there is a conserved
current, ${}^*F$, which could be coupled to a background gauge field
$a_\mu$ as $a\wedge F$.  Supersymmetrizing this coupling, i.e. turning
$a$ into a vector superfield, and giving nonzero expectation value to
the scalar in the $a$ multiplet, we find the FI term.  Therefore, for
every gauge U(1) factor there is a global U(1) symmetry which is not
manifest in the Lagrangian.  The FI term can be thought of as background
vector superfield coupled to the corresponding conserved current.  If
there are $r$ $U(1)$ factors, the above $U(1)^r$ global symmetry
is the Cartan part of, a generally non-Abelian, hidden global symmetry
of rank $r$.

Concretely, we consider gauge theories constructed by Kronheimer
\kron, which are based on the Dynkin diagrams of the $ADE$ gauge
groups.  The Higgs branch gives the ALE spaces \kron.  We argue that
the Coulomb branch in three dimensions gives the moduli space of a small
$ADE$ instanton.  This gives a new connection between the ALE spaces
and the corresponding $ADE$ gauge group.  At the point where the Higgs
and Coulomb branches intersect, we argue that there is an interacting
fixed point.  For the $A_{n-1}$ and $D_n$ cases, we argue that the
theories are dual to the $U(1)$ and $SU(2)$ theories with $n$
hypermultiplets discussed in \refs{\ns, \sw}.  The theories based on
$E_{6,7,8}$ lead to new interesting fixed points which we identify.  The
duality shows that ALE gravitational instantons and the moduli space
of $ADE$ instantons are naturally dual to each other: the dual theories
have one as the Higgs branch and the other has the Coulomb branch.

It is worth stressing that the $E_{6,7,8}$ fixed points give a local
field theory realization of the small $E_{6,7,8}$ instantons, with the
corresponding $E_{6,7,8}$ global symmetry arising as a hidden
symmetry.  In $d=4,5,6$ dimensions it is not known if small $E_n$
instantons correspond to local field theories.  Going to three
dimensions, we exhibit local field theories which flow to these
interacting theories.

The relation between our duality and mirror symmetry of Calabi-Yau
spaces becomes obvious in the context of string compactification.
Consider a pair of mirror Calabi-Yau spaces $\CM$ and $\CM'$.  The map
between them exchanges vector multiplets and hypermultiplets in four
dimensions.  Compactifying further to three dimensions on $\CM \times
S^1$ and $\CM' \times S^1$, the two theories become indistinguishable.
The precise relation between them is that the two radii are inverse of
each other.  Then, the duality discussed in this note becomes identical
to mirror symmetry on the Calabi-Yau space. 

In the next section we introduce and discuss the gauge groups
associated with Kronheimer's ``hyper-Kahler quotient'' of the ALE
spaces \kron.  In sect.\ 3 we discuss the duality between the $A_{n-1}$
case of these theories and $U(1)$ with $n$ electrons.  In sect.\ 4 we
discuss the duality between the $D_{n}$ case and $SU(2)$ with $n$
quarks.  In sect.\ 5 we discuss the $E_{6,7,8}$ cases.

\newsec{The $K_G$ gauge theories}

We consider $N=4$ gauge theories in three dimensions which are naturally
associated with a group $G$ which, for the moment, we take to be simply
laced.  The gauge group is $K_G\equiv (\prod _{i=0}^rU(n_i))/U(1)$,
where $i$ runs over the nodes of the extended Dynkin diagram of the
group $G$ of rank $r$ and $n_i$ is the Dynkin index of the node, with
$n_0=1$ corresponding to the extended node.  The overall $U(1)$ which is
not gauged is the sum of the $U(1)$ generators over all $U(n_i)$.  For
the matter content, we take $\oplus _{ij}a_{ij}({\bf n_i,n_j})$, where
$a_{ij}$ is one if there is a link in the extended Dynkin diagram
connecting nodes $i$ and $j$ and zero otherwise.  Note {}from inspection
of the Dynkin diagrams that every $U(n_i)$ factor has $2n_i$ fundamental
flavors.

These gauge theories were introduced by Kronheimer \kron\ in his
``hyper-Kahler quotient'' construction of the ALE spaces.  In physical
terms: these theories have a Higgs branch on which the gauge group is
completely broken, with one hypermultiplet left massless.  As
discussed in \kron, this Higgs branch is the ALE space $C^2/\Gamma
_G$, with $\Gamma _G$ the discrete $SU(2)$ subgroup
corresponding\foot{For example, $|\Gamma _G|=\sum _i n_i^2$, the nodes
$i$ of the extended Dynkin diagram correspond to the irreducible
representations $R_i$ of $\Gamma _G$, with $|R_i|=n_i$, and $R_F\times
R_i=\sum _ja_{ij}R_j$, with $R_F$ the fundamental two dimensional
representation and $a_{ij}$ as above.} to gauge group $G$.  

The orbifold singularity in $C^2/\Gamma _G$ can be resolved by
introducing Fayet-Iliopoulos $D$ terms in the $K_G$ gauge theory,
corresponding directly to the blowing up modes of \kron.  The $r$ FI
parameters\foot{Each $\vec\zeta _a$ is a real triple, which transforms
in the adjoint of $SU(2)_R$.} $\vec
\zeta _a$, $a=1\dots r$, are naturally considered as being in the
Cartan subalgebra of $G$: turning on some $\vec\zeta _a$ blows up
$C^2/\Gamma _G$ to $C^2/\Gamma _H$, with $H$ related to $G$ by adjoint
Higgsing.  In fact, putting the $IIA$ theory on this space, $G$ is
promoted to a gauge symmetry and the $\vec\zeta _a$ {\it really are}
the flat-directions of a $G$ adjoint matter field.  In the $K_G$ gauge
theory, the added $\vec \zeta _a$ lead to matter expectation values
which Higgs $K_G$ down to the corresponding $K_H$.

The blown up ALE spaces can be written as surfaces in $C^3$ given
by\foot{When we write such expressions, two components of $\vec \zeta$
have been combined into a complex parameter $\zeta$ and the third has
been set to zero.}\ref\kmbu{S. Katz and D.R. Morrison,
Jour. Alg. Geom. {\bf 1} (1992) 449 and references therein.}:
\eqn\aleblow{\eqalign{G&=SU(r+1)\cr G&=SO(2r)\cr G&=E_6\cr G&=E_7\cr
G&=E_8}\quad\eqalign{X_1^2+X_2^2+X_3^{r+1}&=B_G(\zeta, X)\cr
X_1^2+X_2^2X_3+X_3^{r-1}&=B_G(\zeta,X)\cr
X_1^2+X_2^3+X_3^4&=B_G(\zeta,X)\cr
X_1^2+X_2^3+X_2X_3^3&=B_G(\zeta,X)\cr
X_1^2+X_2^3+X_3^5&=B_G(\zeta,X).\cr}} The blowing up polynomials are
$B_G=\sum _{a=1}^rP_{c_a(G)}(\zeta)R_{C_2(G)-c_a(G)}(X)$, where the
subscripts are the degrees of the polynomials under the scaling where
the $\zeta _a$ have degree one and
\aleblow\ has degree $C_2(G)$, the dual Coxeter number of $G$.
$R_{C_2(G)-c_a(G)}(X)$ are the non-trivial chiral ring deformations of
the LHS of \aleblow\ and $c_a(G)$ are the degrees of the Casimirs of
$G$.  The Casimir dependence on $\vec \zeta _i$ again reflects that
they are in the CSA of $G$.

The above description of the Higgs branch of the $K_G$ theories holds
in any dimension upon reduction {}from $N=1$ in six dimensions.  As
discussed in the introduction, the Higgs branch is uncorrected by
quantum effects.  The hyper-Kahler structure completely fixes the
metrics on the spaces \aleblow.

In three dimensions, the above gauge theories have a moduli space of
vacua consisting of two branches: the one\foot{We count dimensions in
quaternionic units, corresponding to four real scalars each.}
dimensional Higgs branch described above and a rank$(K_G)=C_2(G)-1$,
where $C_2(G)$ is the dual Coxeter number of $G$, dimensional Coulomb
branch.  These two branches intersect at a point, where we claim there
is a non-trivial renormalization group fixed point.  At the fixed
point, there is an accidental $G$ global symmetry which is visible only
at long distance.

Unlike the Higgs branch, the Coulomb branch {\it is} corrected by
quantum effects.  Classically, the Coulomb branch is $({\bf R^3}\times
S^1)^{C_2(G)-1}$.  We argue that quantum effects correct it to be the
moduli space of a $G$ instanton, with the point at the origin
corresponding to an instanton of zero size.  Note that the Coulomb
branch has the right dimension, $C_2(G)-1$, to be the moduli space of a
$G$ instanton (eliminating $R^4$ translations).  Further, along the
Coulomb branch the accidental global $G$ symmetry is broken exactly
corresponding to the breaking of $G$ by a $G$ instanton. Turning on
non-zero $\vec \zeta _i$ lifts components of the Coulomb branch
corresponding to the adjoint breaking $G\rightarrow H$: the remaining
components of the Coulomb branch correspond to $K_{H}$.

In addition to the possibility of turning on the FI parameters, we can
consider turning on masses for the hypermultiplets appearing above.
However, a linear combination of the masses can be eliminated for
every $U(1)$ factor in $K_G$ by shifting the origin of the Coulomb
branch.  For $G=A_r$ there are $r+1$ hypermultiplets, associated with
the $r+1$ links of the extended Dynkin diagrams, and $r$ different
$U(1)$ factors, so there is a single mass parameter\foot{$\vec m$ is a
real triple, transforming in the adjoint of $SU(2)_L$.} $\vec m$ which
can not be eliminated.  For $\vec m\neq 0$ the Higgs branch is lifted.
For $G=D_r$ or $G=E_r$ there are $r$ hypermultiplets and $r$ different
$U(1)$ factors, so all masses can be set to zero in these cases.

To summarize, the dimension $d_H$ of the Higgs branch, the
dimension $d_C$ of the Coulomb branch, the number $\#\vec m$ of
possible mass terms, and the number $\#\vec \zeta$ of possible FI terms
are given by:
\thicksize=1pt
\vskip12pt
\begintable
\tstrut | $d_H$ | $d_C$ | $\#\vec m$ | $\#\vec \zeta$ \crthick
$K_{A_{n-1}}$ | $1$ | $n-1$ | $1$ | $n-1$ \cr
$K_{D_n}$ | $1$ | $2n-3$ | $0$ | $n$ \cr
$K_{E_6}$ | $1$ | $11$ | $0$ | $6$ \cr
$K_{E_7}$ | $1$ | $17$ | $0$ | $7$ \cr
$K_{E_8}$ | $1$ | $29$ | $0$ | $8$
\endtable
\noindent
The corresponding quantities for $U(1)$ with $n$ electrons or $SU(2)$
with $n$ quarks are 
\thicksize=1pt
\vskip12pt
\begintable
\tstrut | $d_H$ | $d_C$ | $\#\vec m$ | $\#\vec \zeta$ \crthick
$U(1)$ | $n-1$ | $1$ | $n-1$ | $1$\cr
$SU(2)$ | $2n-3$ | $1$ | $n$ | $0$
\endtable

We now discuss the situation for various $G$ in more detail.

\newsec{$G=SU(n)$}
\subsec{$G=SU(2)$ -- a self-dual example}

The gauge theory $K_{SU(2)}$ constructed as in the previous section is
$U(1)$ with $n=2$ electrons.  The moduli space is a one dimensional
Higgs branch and a one dimensional Coulomb branch, intersecting at the
origin.  It was argued in \ns\ that the theory at the origin is a
non-trivial fixed point.  There are two types of coupling constants
which can be added: a mass difference $\vec m=\vec m_1-\vec m_2$ and a
Fayet-Iliopoulos $D$ term $\vec
\zeta$.  For $\vec m\neq 0$ and $\vec \zeta =0$, 
the $SU(2)_F$ flavor symmetry which rotates the two electrons is
explicitly broken, the Higgs branch is lifted, and there is no
non-trivial fixed point.  For $\vec \zeta
\neq 0$ and $\vec m=0$, the $SU(2)_F$ flavor symmetry is spontaneously
broken, the Coulomb branch is lifted, and there is no non-trivial
fixed point.  For both $\vec m$ and $\vec \zeta\neq 0$, the
moduli space is zero dimensional.

The structure of the Higgs branch is exact at the classical level and
is given by \aleblow\ for $G=SU(2)$; the gauge invariant fields
$X_i=E_f\tilde E ^g(\sigma _i)^f_g$, have \aleblow\ as a classical
constraint with $B_{SU(2)}=\zeta ^2$.  The metric on the Higgs branch
is determined by the hyper-Kahler structure to be that of
Eguchi-Hanson:
\eqn\EHmet{ds^2=g^{2}(\vec x)(dt+\vec\omega\cdot d\vec
x)^2+g^{-2}(\vec x)d\vec x\cdot d\vec x,}
where
\eqn\gis{g^{-2}(\vec x)=\sum _{i=1}^{2}{1\over |\vec x-\vec
\zeta _i|},\qquad \vec{\grad {}}(g^{-2})=\vec{\grad {}} 
\times \vec \omega,} 
and $\vec \zeta_1-\vec \zeta _2=\vec \zeta$ and $i$ labels the two
$U(1)$ gauge groups before we mod out by their sum. See \egh\ for a
review with references on these spaces.

Classically the Coulomb branch is ${\bf R}^3\times S^1$ but, as
discussed in \refs{\ns, \sw}, quantum corrections (essentially given at
one loop)  lead to a hyper-Kahler space with the Taub-NUT metric:
\eqn\TNmet{ds^2=g^{2}(\vec x)(dt+\vec\omega\cdot d\vec
x)^2+g^{-2}(\vec x)d\vec x\cdot d\vec x,} with
\eqn\gis{g^{-2}(\vec x)=g_{cl}^{-2}+\sum _{i=1}^{2}{1\over |\vec x-\vec
m_i|},\qquad \vec{\grad {}}(g^{-2})=\vec{\grad {}}\times \vec \omega,} where
$g_{cl}^{-2}$ is the classical $U(1)$ gauge coupling.  In the limit
$g_{cl}^{-2}=0$, \TNmet\ agrees with \EHmet.

$U(1)$ with $n=2$ electrons has a self duality which exchanges the
classically exact Higgs branch with the purely quantum part of the
Coulomb branch and the FI term $\vec \zeta$ with the mass $\vec m$.
The fact that it is necessary to take $g_{cl}\rightarrow
\infty$ in \TNmet\ is standard for duality in 
theories which are not-finite: duality is a property of the
long-distance physics \nso.  Further, the Higgs (Coulomb) branch is the
moduli space of an $SU(2)$ instanton: the hyper-Kahler $SU(2)_R$
($SU(2)_L$) action corresponds to $SU(2)$ rotations of the instanton
and the distance {}from the origin corresponds to the instanton size.

For $\vec m=\vec \zeta=0$, $SU(2)_F\times SU(2)_R$ is spontaneously
broken along the Higgs branch to a diagonal subgroup.  It is unbroken
along the Coulomb branch.  The duality says that there is a new
$\widetilde{SU(2)}_F $ symmetry at long distance.
$\widetilde{SU(2)}_F\times SU(2)_L$ is spontaneously broken to a
diagonal subgroup along the Coulomb branch but unbroken along the Higgs
branch.

\subsec{$G=SU(n)$, $n>2$}

The gauge theory is $K_{SU(n)}=U(1)^{n}/U(1)\cong U(1)^{n-1}$,
with $n$ hypermultiplets $Q _i$
charged under $U(1)_i$ and $U(1)_{i+1}$ for $i=0\dots n-1$, with
$U(1)_{n}\equiv U(1)_0$.  There is a one dimensional Higgs branch,
which intersects an $n-1$ dimensional Coulomb branch at the origin.
There are $n-1$ independent FI terms: $\vec\zeta _i$, $i=0\dots
n-1$, with $\sum _i\vec\zeta _i=0$, and one independent mass term,
$\vec m$.  For $\vec m\neq 0$, the Higgs branch is lifted.  For $\vec
\zeta _i\neq 0$ some of the Coulomb branch is lifted.

We argue that these theories have non-trivial fixed points where they
are dual to the $U(1)$ theories with $n$ electrons discussed in \ns.
This dual theory has a one dimensional Coulomb branch which intersects
a $n-1$ dimensional Higgs branch at the origin.  The Higgs branch of
the $U(1)^{n-1}$ theory is mapped to the Coulomb branch of
the $U(1)$ dual and visa-versa.  

Following the discussion in the introduction, the $K_{SU(n)}\cong
U(1)^{n-1}$ theory has a hidden global $U(1)^{n-1}$ symmetry which
couples to the FI terms $\vec \zeta _i$.  The duality shows that this
global symmetry is promoted to a global $SU(n)$ symmetry at the fixed
point, which is the manifest flavor symmetry of the dual theory.  Under
the duality, the $n-1$ FI terms of the $K_{SU(n)}$ theory are mapped to
the $n-1$ independent masses of the $U(1)$ dual, which we write as $\vec
m_i'$, $i=0\dots n-1$, with $\sum _i\vec m _i'=0$.  The single mass
$\vec m$ of the $K_{SU(n)}$ theory is mapped to the FI parameter
$\vec\zeta '$ of the $U(1)$ theory.

Turning on $\vec \zeta _i$ FI terms leads to hypermultiplet
expectation values which Higgs $K_{SU(n)}\rightarrow K_H$, with $H$
related to $SU(n)$ by adjoint breaking.  In the dual $U(1)$ theory,
this corresponds to added mass terms, which leads to the dual of
$K_{H}$. For example, taking $\vec
\zeta _i\neq 0$ with $n-1$ equal values breaks
$U(1)^{n}/U(1)\rightarrow U(1)^{n-1}/U(1)$.  This is mapped to
giving an electron a mass in the $U(1)$ dual, leaving $n-1$ light
flavors.  The duality is thus preserved in the low energy theory.
Similarly, for $\vec m\neq 0$ in the $K_{SU(n)}$ theory the Higgs
branch is lifted, which is mapped to the lifting of the Coulomb branch
for $\vec \zeta '\neq 0$ in the $U(1)$ dual.

The Higgs branch of $U(1)$ with $n$ electrons was interpreted in
\mrd, as the moduli space of $SU(n)$ instantons (modulo
$R^4$ translations).  This is consistent with the fact that the global
$SU(n)$ symmetry is broken to $SU(n-2)\times U(1)$ on the Higgs
branch with light hypermultiplets transforming as $({\bf
n-2})_1\oplus {\bf 1}_0$ \swoii.  In the dual theory, the hidden
global $SU(n)$ theory is so broken along the Coulomb branch.

As further evidence for the duality, we now compare the Higgs and the
Coulomb metrics in the dual theories, showing that they are
interchanged.

The Higgs branch of the $K_{SU(n)}$ theory is given exactly by the
classical result in \aleblow, with metric given by
\eqn\analemet{ds^2=g^{2}(\vec x)(dt+\vec\omega\cdot d\vec
x)^2+g(\vec x)^{-2}d\vec x\cdot d\vec x,}
with 
\eqn\giss{g^{-2}(\vec x)=
\sum _{i=0}^{n-1}{1\over |\vec x-\vec \zeta _i|},\qquad 
\vec{\grad {}} (g^{-2})=\vec{\grad {}}
\times \vec \omega.}  

Classically the Coulomb branch of the $K_{SU(n)}$ theory is $({\bf
R^3}\times S^1)^{n-1}$.  Quantum mechanically, the metric is
corrected to be the multi-dimensional version of Taub-NUT:
\eqn\multidtn{ds^2=(g^{-2})_{ij}d\vec x_i\cdot d\vec x_j
+(g^2)_{ij}dq_idq_j,}
with $dq_i=dt_i+\vec W_{ij}\cdot d\vec x_j$ and
\eqn\multigis{\eqalign{(g^{-2})_{ii}&=
g_{i,cl}^{-2}+\sum _{j\neq i}{a_{ij}\over |\vec x_i-\vec x_j-\vec
m_i|},\cr (g^{-2})_{ij}&=-{a_{ij}\over |\vec x_i-\vec x_j -\vec
m_i|},\qquad i\neq j,}} with $a_{ij}$ the Dynkin diagram adjancency
matrix and $\vec{\grad i}(g^{-2})_{ij}=\vec{\grad i}\times \vec W_{ij}$.  This
metric coincides with the multi-monopole metrics found, for example,
in \ref\multimon{K. Lee, E. Weinberg, P. Yi, ``The Moduli Space of
Many BPS Monopoles for Arbitrary Gauge Groups,'' hep-th/9602167.}.  In
the present context, this metric is obtained by essentially the same
one-loop calculation which entered in the $n=2$ case \gis: each
$U(1)_i$ has precisely two electrons carrying its charge and the
electrons get an additional effective mass contribution {}from the
coupling to another gauge field.  As mentioned before, by shifting the
$\vec x_i$, the masses $\vec m_i$ can be eliminated up to a single
linear combination $\vec m$.  

The Higgs branch of $U(1)$ with $n$ electrons is given exactly by
the classical result: the metric the same as \multidtn\ with
$g_{i,cl}^{-2}\rightarrow 0$ and $\vec m_i\rightarrow \vec \zeta '_i$,
where the FI term $\vec \zeta '$ is related to the $\vec \zeta _i'$ so
that $\vec m\rightarrow \vec \zeta '$.

The Coulomb branch of $U(1)$ with $n$ electrons is classically ${\bf
R^3}\times S^1$ but, owing to quantum corrections, is given by
\eqn\anmet{ds^2=g^{2}(\vec x)(dt+\vec\omega\cdot d\vec
x)^2+g(\vec x)^{-2}d\vec x\cdot d\vec x,} with
\eqn\gis{g^{-2}(\vec x)={g_{cl}'}^{-2}
+\sum _{i=0}^{n-1}{1\over |\vec x-\vec m_i'|},\qquad \vec{\grad {}}
(g^{-2})=\vec{\grad {}}  \times \vec \omega.}  
This Coulomb branch metric coincides with the Higgs branch metric
\analemet\ upon taking ${g_{cl}'}^{-2}=0$ and replacing $\vec \zeta
_i\rightarrow \vec m_i'$.

\newsec{$G=SO(2n)$}

The gauge group is $K_{SO(2n)}=U(1)^4\times U(2)^{n-3}/U(1)\cong
U(1)^3\times U(2)^{n-3}$.  Corresponding to the links in the extended
$SO(2n)$ Dynkin diagram, there are two doublets under the first $U(2)$
and two under the last $U(2)$, each charged under a $U(1)$, and matter
fields transforming as $({\bf 2,2})$ under each adjacent $U(2)\times
U(2)$.  We argue that these theories have non-trivial fixed points, and
that they are dual to the $SU(2)$ with $n$ quark flavors discussed in
\ns.  Under the duality, the one dimensional Higgs branch of the
$K_{SO(2n)}$ theory is mapped to the one dimensional Coulomb branch of
the $SU(2)$ dual and the $2n-3$ dimensional Coulomb branch of the
$K_{SO(2n)}$ theory is mapped to the $2n-3$ dimensional Higgs branch
of $SU(2)$ with $n$ quarks.  At the fixed point, the $K_{SO(2n)}$
theory has a hidden global $SO(2n)$ symmetry which is mapped to the
manifest $SO(2n)$ of the $SU(2)$ dual.  The $n$ independent FI
deformations of the $K_{SO(2n)}$ theory are mapped to the $n$
independent masses for the quark flavors of the $SU(2)$ dual.  The
fact that the $SU(2)$ theory has no FI term deformation corresponds to
the fact, noted in sect. 2, that the $K_{SO(2n)}$ theory has no
hypermultiplet mass term deformation.

Turning on FI terms in the $K_{SO(2n)}$ theory leads to matter field
expectation values which break $K_{SO(2n)}$ to $K_{H}$, with $H$
related to $G$ by adjoint Higgsing, spontaneously breaking the hidden
global $SO(2n)$ symmetry to a hidden global $H$ symmetry.  In the
dual, the corresponding mass terms explicitly break the manifest
$SO(2n)$ symmetry to $H$.  As a particular example, turning on equal
FI parameters breaks $K_{SO(2n)}$ to $K_{SU(n)}$, with a remaining
$SU(n)$ global symmetry.  In the $SU(2)$ dual, the common mass for the
$n$ quarks leads to a vacuum with $SU(2)$ Higgsed to $U(1)$ with $n$
massless electrons.  The low energy theories are again dual.

The Higgs branch of the $K_{SO(2n)}$ theories is given by the classical
result \aleblow.  The Coulomb branch of $SU(2)$ with $n$ quarks receives
quantum corrections, becoming the same $D_{n}$ ALE space \ns\ in the
$g_{cl}\rightarrow \infty$ limit.  Similarly, the quantum Coulomb branch
of the $K_{SO(2n)}$ theory is expected to coincide with the Higgs
branch of $SU(2)$ with $n$ quarks.

The Coulomb branch of the $K_{SO(2n)}$ theory or the Higgs branch of
the $SU(2)$ theory gives the $SO(2n)$ instanton moduli.  This is
compatible with the fact that the global symmetry in the $SU(2)$
theory is broken as $SO(2n)\rightarrow SO(2n-4)\times SU(2)$ on this
space with massless hypermultiplets $\half ({\bf 2n-4,2})\oplus ({\bf
1,1})$ \swoii.  In the $K_{SO(2n)}$ theory the hidden global $SO(2n)$
symmetry must be so broken along the Coulomb branch.  The Cartan part
of that can be seen as in the introduction.

\newsec{$G=E_{6,7,8}$}

For these cases, we argue that the $K_G$ gauge theory again leads to a
non-trivial fixed point, with a Coulomb branch which gives the moduli
space for $G$ instantons.  In these cases, it is not known what the
dual theories are whose Higgs branch is the Coulomb branch of the
$K_G$ theory.  Again, we stress that the $K_{E_{6,7,8}}$ gives a local
field theory description of the interesting phenomenon associated with
small $E_{6,7,8}$ instantons in string theory.

For $E_6$, we want the Coulomb branch to reflect the global symmetry
breaking $E_6\rightarrow SU(6)$, with the massless Coulomb moduli
transforming like $\half ({\bf 20})\oplus {\bf 1}$.  The dimension
agrees with the rank, $C_2(E_6)-1$, of $K_{E_6}$.  Again, there should
be a hidden global $E_6$ symmetry at the origin which is so broken
along the Coulomb branch.  As explained above, its Cartan subgroup is
dual to the $U(1)^6$ gauge symmetry.

The situation for $E_7$ and $E_8$ is similar.  For $E_7$, we want the
Coulomb branch to reflect the global symmetry 
breaking $E_7\rightarrow SO(12)$, with the massless Coulomb moduli
transforming like $\half ({\bf 32})\oplus {\bf 1}$.
For $E_8$, we want the Coulomb branch to reflect the global symmetry
breaking $E_8\rightarrow E_7$, with the massless Coulomb moduli
transforming like $\half ({\bf 56})\oplus {\bf 1}$.  

\bigskip
\centerline{{\bf Acknowledgments}}

We would like to thank G. Moore, S. Shenker, and E. Witten for
discussions.  KI thanks the Aspen Center for Physics, where this work
was partially completed.  This work was supported in part by DOE grant
\#DE-FG02-96ER40559, NSF grant PHY-9513835, and the W.M. Keck
Foundation.

\midinsert
\centerline{\hglue1.5in\psfig{figure=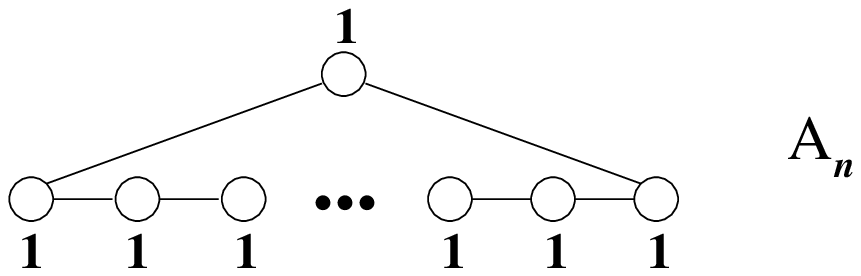,width=4.3in,angle=0}}
\bigskip
\centerline{\psfig{figure=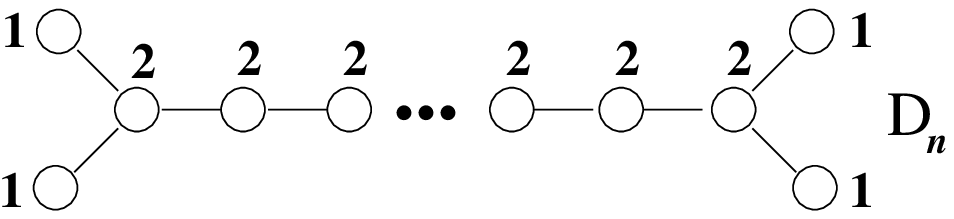,width=3.8in,angle=0}}
\bigskip
\centerline{\psfig{figure=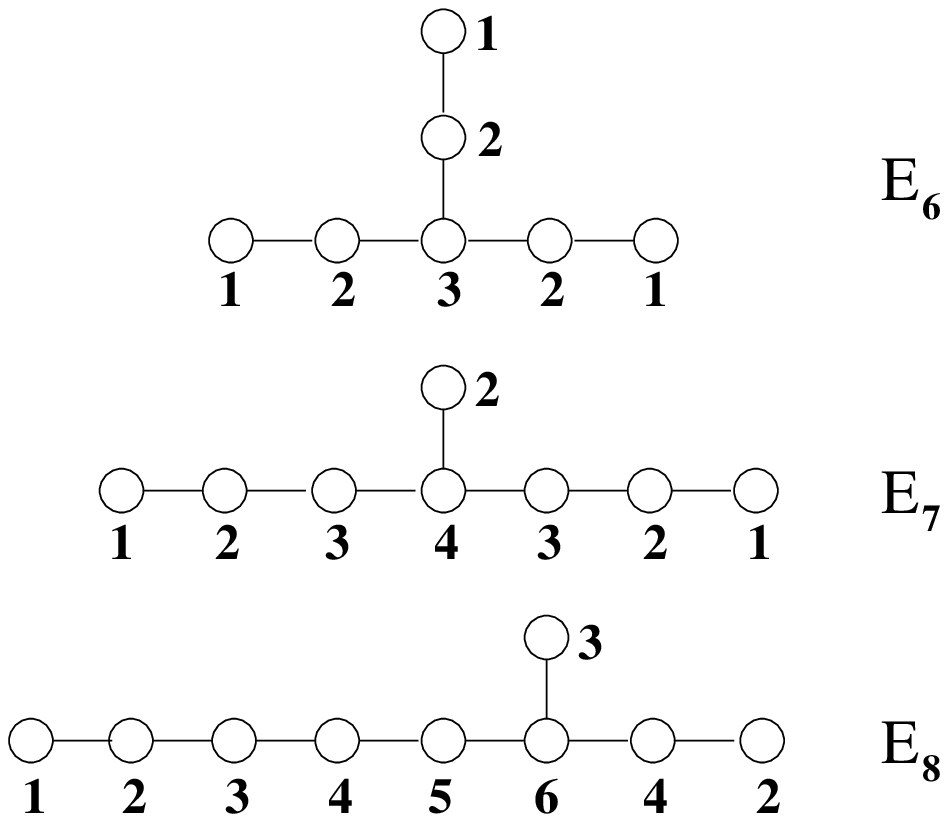,width=3.8in,angle=0}}
Extended Dynkin diagrams and indices.
\endinsert
\listrefs
\end